%% file: skeleton.tex
\documentclass[a4paper,11pt]{article}
\usepackage{pos}
\usepackage{comment}
\input{newcommands}

\title{Inflation driven by repulsive-like primordial black holes}

\author[a,b]{Konstantinos Dialektopoulos}

\author*[c,d,e]{Theodoros Papanikolaou}

\author[f]{Vasilios Zarikas}

\affiliation[a]{Department of Mathematics and Computer Science,
Transilvania University of Brasov, 500091, Brasov, Romania}
\affiliation[b]{Institute of Space Sciences and Astronomy, University of Malta, Msida, Malta and Department of Physics, University of Malta, Msida, Malta}

\affiliation[c]{Laboratory of Theoretical and Computational Physics, Department of Physics, University of Patras, 26504 Patras, Greece}

\affiliation[d]{Scuola Superiore Meridionale, Via Mezzocannone 4, 80134 Napoli, Italy}

\affiliation[e]{Istituto Nazionale di Fisica Nucleare (INFN), Sezione di Napoli, Via Cinthia 21, 80126 Napoli, Italy}

\affiliation[f]{Department of Mathematics, University of Thessaly, 35100, Lamia, Greece}

\emailAdd{papaniko@upatras.gr}

\abstract{We review a new natural inflationary mechanism operated by repulsive-like primordial black holes (PBHs). In particular, working within the ``Swiss Cheese" cosmological framework, we find that a Universe filled with PBHs, whose spacetime metric presents a repulsive-like behaviour, is characterised by an early quasi-de-Sitter cosmic expansion phase. Notably, for light PBHs with \( m < 5 \times 10^8 \) g, evaporating before Big Bang Nucleosynthesis (BBN), one is met with an exponential inflationary phase with graceful exit and reheating proceeding through PBH evaporation. Furthermore, one finds as well that PBHs with \( m \sim 10^{12} \) g and abundances \( 0.107 < \Omega^\mathrm{eq}_\mathrm{PBH} < 0.5 \) near matter-radiation equality can act as an early dark energy component, easing in this way naturally the Hubble tension.}

\FullConference{Proceedings of the Corfu Summer Institute 2025 "School and Workshops on Elementary Particle Physics and Gravity" (CORFU2025)\\
27 April - 28 September, 2025\\
Corfu, Greece\\}

\begin{document}
\maketitle

\medskip
\section{Introduction}
Primordial black holes (PBHs), first introduced back in 1970s~\cite{1967SvA....10..602Z, Carr:1974nx,1975ApJ...201....1C,1979A&A....80..104N}, can be produced through a plethora of formation mechanisms, such as the collapse of enhanced cosmological perturbations~\cite{Carr:1975qj,Musco:2020jjb}, topological defects~\cite{Hawking:1987bn,Polnarev:1988dh}, primordial phase transitions~\cite{Hawking:1982ga,Moss:1994iq,Jung:2021mku} and modified or quantum gravity scenarios~\cite{Barrow:1996jk,Kawai:2021edk,Papanikolaou:2023crz}.  After the recent discovery of the first GWs from binary mergers in 2015~\cite{LIGOScientific:2016aoc}, PBHs have rekindled the interest of the scientific community since they can explain some of the recent black hole merging evening detected by LIGO/VIRGO~\cite{Sasaki:2016jop,Franciolini:2021tla}. They can also account for a part or the totality of dark matter~\cite{Chapline:1975ojl,Carr:2020xqk} 
as well as constituting potential seeds of large scale structures (LSS)~\cite{Meszaros:1975ef,Afshordi:2003zb}. Furthermore, PBHs can drive reheating through their evaporation~\cite{Lennon:2017tqq}, playing as well an important role on baryogenesis~\cite{Barrow:1990he, Baumann:2007yr, Klipfel:2026nzx},  supermassive black hole generation~\cite{Ziparo:2024nwh,Prole:2025snf,Zhang:2025grn,Haque:2026vvp,Kallifatides:2026sik} as well as on the alleviation of the Hubble tension~\cite{Papanikolaou:2023oxq}. For recent PBH reviews, the interested reader can study~\cite{Carr:2020gox,Escriva:2022duf}.

A standard assumption in PBH studies is that the PBH spacetime is described by the Schwarz-schild or the Kerr metric~\cite{Khlopov:2008qy,Carr:2016drx,Escriva:2022duf,LISACosmologyWorkingGroup:2023njw,Choudhury:2024aji} which are typically non-singular in the center and attractive. In this work, we relax this assumption by focusing on PBHs which can repel test mass particles either on large or small distances from the horizon and which are mostly related to regular spacetimes~\cite{Luongo:2023aib}. Interestingly enough such regular PBHs can address at the same time the dark matter and the singularity problems~\cite{Easson:2002tg,Dymnikova:2015yma,Pacheco:2018mvs,Arbey:2021mbl,Arbey:2022mcd,Banerjee:2024sao,Davies:2024ysj,Calza:2024fzo,Calza:2024xdh}. For review paper on regular black hole spacetimes, one can read~\cite{Ansoldi:2008jw,Nicolini:2008aj,Torres:2022twv,Lan:2023cvz}.

In particular, we will review in this work a novel inflationary mechanism driven by PBHs with repulsive behaviour~\cite{Dialektopoulos:2025mfz}. In order to see how a population of PBHs can influence collectively cosmic expansion, we apply  a ``Swiss - Cheese'' cosmological framework, firstly conceptualized by Einstein and Strauss in $1940s$~\cite{Einstein:1945id}. In ``Swiss Cheese'' cosmology, one matches basically consider a homogeneous and isotropic cosmological spacetime as a ``cheese" filled with black holes. To describe thus collectively the dynamical behaviour of such a Universe, one needs to match a black hole spacetime with the cosmological one. For the case of Schwarzschild black holes, one trivially obtains a dust-filled Universe. 

As it will become more clear later, for the case of repulsive-like PBHs, one is inevitably met with an early accelarated cosmic expansion. Astonishingly, this new non-conventional mechanism of cosmic acceleration can lead naturally to an inflationary era with graceful exit, terminated at an energy scale dependent on PBH metric parameters or due to black hole evaporation. It can play as well a role in generating an early dark energy (EDE) component of non scalar-field nature, potentially alleviating the $H_0$ cosmic tension.

\section{The Hayward prototype regular black hole spacetime}

Regular black hole spacetimes can arise in theories of quantum gravity, resolving naturally curvature singularities~\cite{Cadoni:2023nrm, Bonanno:2023rzk, dePaulaNetto:2023cjw,Singh:2009mz,Ashtekar:2023cod}. Within the quantum gravity framework, black holes can be modeled by effective classical spacetimes with high-curvature corrections, being characterized by non-singular cores~\cite{1968qtr..conf...87B,Frolov:1981mz,Roman:1983zza,Dymnikova:1992ux,Borde:1996df,Hayward:2005gi,Hossenfelder:2009fc,Bambi:2013gva,Bambi:2013caa,Hawking:2014tga,Frolov:2014jva,Bardeen:2014uaa,Haggard:2014rza,Barrau:2015uca,Haggard:2015iya}. Below, we illustrate our cosmic acceleration mechanism using the Hayward metric, which will serve as a prototype regular black hole spacetime. The original Hayward metric was originally introduced in $2006$ and can be written effectively as~\cite{Hayward:2005gi}
\begin{equation}\label{eq:Hayward_metric}
ds^2=-F(R)\,dt^2+\frac{1}{F(R)}\,dR^2+R^2\,d\Omega^2 ~,
\end{equation}
where
\begin{equation}\label{hayward}
F(R)=1-\frac{2\,G_\mathrm{N} M(R)}{R}~,
\end{equation}
and
\begin{align}
M(R)=\frac{m\,R^3}{R^3+2\,G_\mathrm{N}\, m\,L^2}  \approx
\begin{cases}
m & (R  \gg G_\mathrm{N}^{1/3} m^{1/3} L^{2/3}) \\
R^3/(2G_\mathrm{N} L^2) & (R  \ll G_\mathrm{N}^{1/3} m^{1/3} L^{2/3}),
\end{cases}
\end{align}
where $m$ stands for the mass of the black hole at asymptotic infinity.

Regarding the horizon structure of the Hayward metric \eqref{eq:Hayward_metric}, one can show that the latter admits two Killing horizons, determined by the roots of
\begin{equation} \label{hor}
F(R) = 1-\frac{2\,G_\mathrm{N}\,m\,R^2}{R^3+2\,L^2\,G_\mathrm{N}\, m}=0.
\end{equation}
In particular, for $m>\frac{3\sqrt{3}L}{4G_\mathrm{N}}$, \Eq{eq:Hayward_metric} admits inner and outer horizons at approximately $R \simeq L$ and $R \simeq 2\,G_\mathrm{N}\,m$, respectively~\cite{Hayward:2005gi,DeLorenzo:2014pta}.

Other regular black holes spacetimes which were explored include the Bardeen and the Dymnikova black holes, being discussed in detail in the Appendices of \cite{Dialektopoulos:2025mfz}. Moreover, the  de Sitter–Schwarzschild spacetime, being singular in the center but repulsive on large scales, was also studied. Interestingly enough, the  de Sitter–Schwarzschild metric can be basically regarded as a limiting case of the McVittie spacetime~\cite{1933MNRAS..93..325M}, approximating according to recent literature quite well the general relativistic PBH metric during the collapse phase~\cite{DeLuca:2020jug,Hutsi:2021vha,Hutsi:2021nvs}.

\section{The emergence of cosmic acceleration}
Having recap above our prototype regular PBH metric, let us now describe collectively the effect of our PBH ``gas", dominating the energy budget of the Universe, on cosmic expansion. Doing so, we apply the ``Swiss-Cheese" cosmological framework by matching the Hayward metric with an exterior homogeneous and isotropic background. More specifically, one basically apply the Darmois-Israel junction conditions by imposing the continuity of the intrinsic metric $\gamma_{\alpha\beta}$ and the extrinsic curvature $K_{\alpha\beta}$ on a spherical $3D$-surface $\Sigma$ which is at fixed coordinate radius in the cosmological frame but evolving in the black hole frame. For more technical details the interesting reader should study the Appendix A of~\cite{Dialektopoulos:2025mfz} as well as~\cite{Kofinas:2017gfv}. At this point, one should clarify as well the fact that we apply the ``Swiss-Cheese'' cosmological framework in the early Universe, where a homegeneously distributed in space PBH population is a very good approximation~\cite{Desjacques:2018wuu, Ali-Haimoud:2018dau, MoradinezhadDizgah:2019wjf,DeLuca:2022uvz}, unlike the present epoch. 

At the end, performing the Darmois-Israel junction matching for a single Hayward black hole, we obtain that the background cosmic expansion dynamics can be recast as:
\begin{equation}\label{eq:H_2_Hayward}
H^2=\frac{\dot{a}^2}{a^2}=\frac{2G_\mathrm{N} m}{R^3+2 G_\mathrm{N} m L^2}-\frac{k}{a^2}\,,  \quad \frac{\ddot{a}}{a}=\frac{G_\mathrm{N} m (4\,G_\mathrm{N} L^2 m-R^3)}{(2 G_\mathrm{N} L^2 m+R^3)^2}\,, 
\end{equation}
where $ R = a\,r_{\Sigma} $, $ a $ denotes the scale factor, $ H=\dot{a}/a $  the Hubble parameter and $\,r_{\Sigma} $ the comoving Shucking radius emerging out of the ``Swiss-Cheese" cosmological analysis~[See ~\cite{Dialektopoulos:2025mfz} for more details.].

One thus can derive the overall effect of a collective population of PBHs by modeling our Universe as a ``Swiss-Cheese" with each spherical spherical vacuole filled with a PBH. Considering thus a PBH-dominated Universe we identify the energy density in the exterior region of a black hole with its interior mass density, i.e. 
\beq\label{eq:rho_vs_a}
\rho = \frac{3m}{4\pi(a r_{\Sigma})^3},
\eeq 
Putting thus \eqref{eq:rho_vs_a} into the second Eq. of~\eqref{eq:H_2_Hayward}, the background cosmic expansion equations read as:
\begin{equation}
\frac{\ddot{a}}{a}= 4 \pi G_\mathrm{N} \rho \,  \frac{(16 G_\mathrm{N} L^2 \pi \rho - 3)}{(8 G_\mathrm{N} L^2 \pi \rho + 3)^2}.
\label{acceleration with density}
\end{equation}
As one can clearly see from \Eq{acceleration with density}, $\ddot{a}/a$ changes sign at a critical density $\rho_{\rm c}$ given by
\begin{equation}\label{rhoch}
\rho_{\rm c}=\frac{3}{16 \pi G_\mathrm{N}L^2}.
\end{equation}
Namely, for $\rho > \rho_{\rm c}$, the Universe experiences an accelerated cosmic expansion phase, while for $\rho < \rho_{\rm c}$ one gets a standard cosmic deceleration era. Inflation thus terminates naturally, without any fine-tuning.


\section{Cosmological consequences of a PBH-driven inflationary era}

Let us now see two interesting cosmological implications of our novel PBH-driven inflationary scenario.

\subsection{Inflation with graceful exit and reheating}  
An early PBH-dominated Universe constitutes a plausible scenario~\cite{Nagatani:1998gv,Conzinu:2023fui}, in particular within the context of quantum gravity realized PBH spacetimes~\cite{DeLorenzo:2014pta, Shafiee:2022jfx, Bonanno:2020fgp}. Let us then consider a ``gas" of randomly distributed PBHs of Hayward type, which dominate the energy budget of the Universe before BBN~\footnote{One should stress here that even within the standard Hot Big Bang (HBB) cosmology, PBHs forming during a radiation-dominated (RD) era will dominate ultimately the Universe energy density due to the slower dilution of their density compared to that of radiation~\cite{Hidalgo:2011fj,Suyama:2014vga,Zagorac:2019ekv,Hooper:2019gtx}.}. For simplicity, we will assume that all PBHs have the same mass.

In a spatially flat Universe ($k=0$), the modified cosmic dynamics~\eqref{eq:H_2_Hayward} will read as
\begin{equation}
H^2=\frac{8\pi}{3}G_\mathrm{N}\left( \rho^{-1} +\frac{8\pi}{3}G_\mathrm{N}L^2\right)^{-1}.
\label{modH^2}
\end{equation}
At high densities, $H^2 \simeq L^{-2}$, and one gets inevitably a quasi-de Sitter inflationary phase with a nearly exponentially growing scale factor $
a(t) \propto e^{t/L}$. Inflation terminates then when either PBHs evaporate or, as one can see from Eq.~\eqref{modH^2}, when $\rho$ drops below $\rho_{\rm e} \simeq G_\mathrm{N}^{-1}L^{-2}$. In the post-inflationary phase however, if PBH evaporation has not yet been completed, the cosmic dynamics will be governed by the following equation:
\begin{equation}
H^2 = \frac{8\pi}{3}G_\mathrm{N} \frac{\rho_{\rm e} a^3_{\rm e}}{a^3},
\end{equation}
since in this regime, $\rho \simeq \rho_{\rm BH} = \rho_{\rm e} a^3_{\rm e} / a^3$, where the index $\mathrm{e}$ denotes the end of inflation and $\rho_{\rm BH}$ stands for the PBH mass density. PBHs will then evaporate à la Hawking ~\cite{Hawking:1974rv}, with their evaporation being completed either during or after the quasi-de Sitter phase, leading to a graceful inflationary exit and a natural transition to the RD era where $H^2 \propto a^{-4}$.

At this point, it is important to make a comment as well regarding the perturbative behaviour of our inflationary scenario. In particular, one should emphasize that given the quasi-de Sitter cosmic expansion, the first and second Hubble flow slow-roll parameters $\epsilon_1$ and $\epsilon_2$ defined as $\epsilon_1 \equiv -\dot{H}/H^2$ and $\epsilon_2 \equiv \frac{1}{\epsilon_1}\frac{\mathrm{d}\epsilon_1}{\mathrm{d}N}$ will be inevitably very small, leading to a spectral index $n_\mathrm{s} \simeq 1 - 2\epsilon_1 - \epsilon_2 \simeq 1$ and a tensor-to-scalar ratio $r\simeq 16\epsilon_1 \ll 1$ during inflation, being compatible with the CMB data~\cite{Planck:2018vyg}.

Concerning now reheating, the latter will proceed naturally through Hawking PBH evaporation. Making use of the time-dependent Hayward metric~\cite{Frolov:2017rjz}:
\begin{equation}
\label{Hayward}
F = 1 - \frac{2 G_\mathrm{N} m(t) R^2}{R^3 + 2 G_\mathrm{N} m(t) L^2},
\end{equation} 
the Hawking law of PBH mass loss will be written as
\begin{equation}
\label{evap}
\frac{dm(t)}{dt} \sim - \frac{1}{C^3 G_\mathrm{N}^2 m(t)^2},
\end{equation}
where $ C = \left(\frac{640\pi}{n_{\rm p}}\right)^{1/3} \left(\frac{L}{l_\mathrm{Pl}}\right)^{2/3} $ and $n_{\rm p}$ is the number of particle polarizations. Integrating \Eq{evap} the PBH lifetime will be recast as
\begin{equation}
\label{firstconstraint}
t_{\rm evap} = \frac{1}{3} C^3 G_\mathrm{N}^2 m^3.
\end{equation}
In order thus for evaporation to terminate before BBN ($t_{\rm BBN} \sim 1\, \text{min}$), we require that:
\begin{equation}
\label{constr1}
t_{\rm evap} \leq t_{\rm BBN} \Rightarrow m \leq G_\mathrm{N}^{-2/3} C^{-1} (3 t_{\rm BBN})^{1/3}.
\end{equation}
For $n_{\rm p} \sim 100$ and $L = 100\,l_{\rm Pl}$, we obtain trivially that $m < 5 \times 10^8$ g.
As long as Eq. \eqref{constr1} 
holds, our ``Swiss-Cheese" inspired mechanism ensures a consistent inflationary era with graceful exit and reheating occuring via PBH Hawking evaporation.

A crucial question then to ask is when will be the end of inflation in this alternative scenario. To answer this question, one needs to compare the critical density $\rho_{\rm c}$ \eqref{rhoch} with the energy density at PBH evaporation time $\rho_{\rm evap}$, assuming a RD post-evaporation era. Considering thus Eq.~\eqref{firstconstraint} together with the fact that $H_{\rm evap} = 1/2t_{\rm evap}$, we find that
\begin{equation}
\label{eq:rho_evap}
\rho _{\rm evap} = 27648 \pi ^4 \Mp ^4 \left(\frac{\Mp}{m}\right)^6 \left(\frac{n_{\rm p}}{640\pi}\right)^{2/3} \left(\frac{l _{\rm Pl}}{L}\right)^{4/3},
\end{equation}
where $\Mp ^2 = 1 / (8\pi G_\mathrm{N})$.

\begin{figure*}[ht!]
\centering
\includegraphics[width=0.49\textwidth]{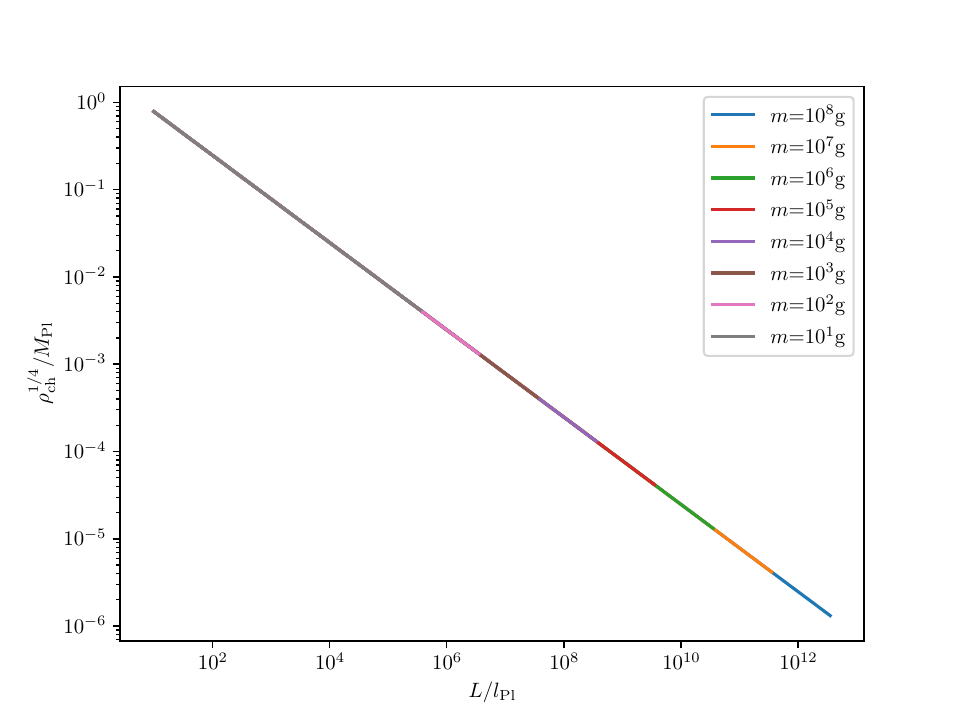}
\includegraphics[width=0.49\textwidth]{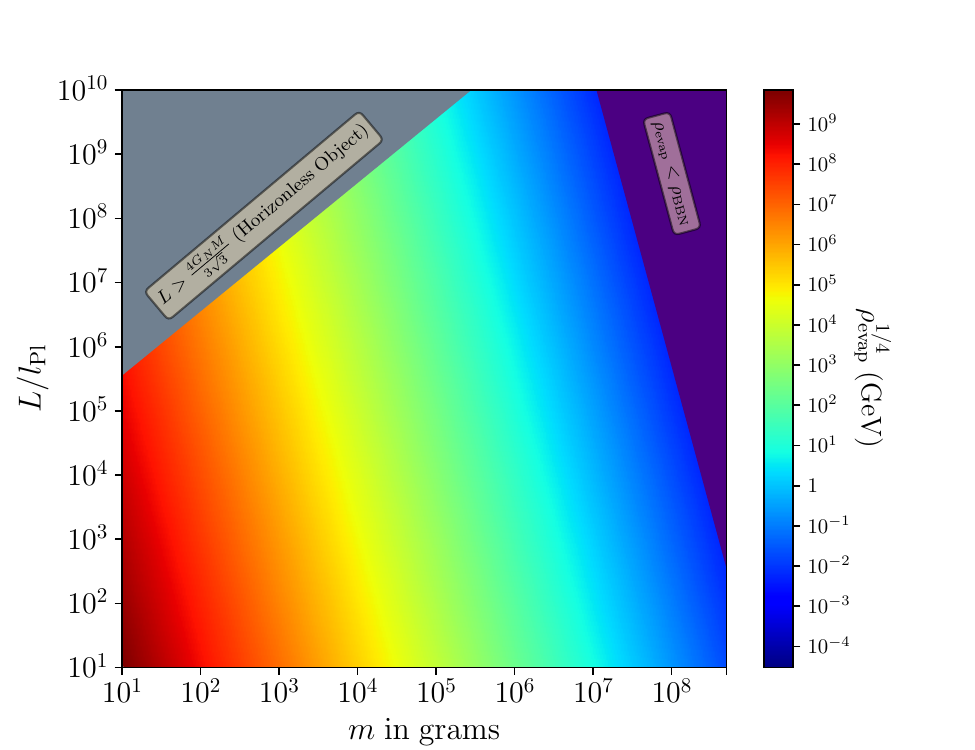}
\caption{{Left Panel: We depict $\rho_\mathrm{c}$ as a function of the regularising fundamental length scale $L$.  Right Panel: We display $\rho_\mathrm{evap}$ (color-bar axis) as a function of the PBH mass $m$ ($x$-axis) and the scale $L$ ($y$-axis). The grey region $L>\frac{4G_\mathrm{N}m}{3\sqrt{3}}$, where horizonless objects are realized, is not of theoretical interest,  while the magenta one, where $\rho_\mathrm{evap}<\rho_\mathrm{BBN}$ is observationally excluded.}}
\label{fig:rho_c_vs_rho_evap}
\end{figure*}

In \Fig{fig:rho_c_vs_rho_evap}, the left panel depicts $\rho_{\rm c}$ as a function $L$, with the region $L < \frac{4G_\mathrm{N}m}{3\sqrt{3}}$ describing horizonfull objects. On the other hand, the right panel shows $\rho_{\rm evap}$ as a function of both $m$ and $L$. The grey region, denoting horizonless objects, is not of theoretical interest whereas the magenta one, associated to PBH evaporation taking place after BBN, is observationally excluded. In particular, the magenta region determined by the condition $\rho_{\rm evap} < \rho_{\rm BBN}$, can be described by
\begin{equation}
L > 7 \times 10^5\, l_\mathrm{Pl} \left( \frac{10\, \text{MeV}}{\rho_\mathrm{BBN}^{1/4}} \right)^3 \left( \frac{n_{\rm p}}{640\pi} \right)^{1/2} \left( \frac{10^8\, \text{g}}{m} \right)^{9/2}.
\end{equation}

As it can be inferred from \Fig{fig:rho_c_vs_rho_evap}, for the case of Hayward PBHs, the inequality $\rho_{\rm c} > \rho_{\rm evap}$ is always valid, with inflation terminating before PBH evaporation. Similar findings were deduced for the Bardeen and Dymnikova spacetimes as well, where $t_{\rm evap} \propto \tilde{C} t_{\rm S}$ with $\tilde{C} > 1$~\cite{Calza:2024fzo}. For the singular Schwarzschild–de Sitter case, inflation always ends due to PBH evaporation [See \cite{Dialektopoulos:2025mfz} for more details].

\subsection{Early dark energy contribution before matter-radiation equality}
Let us now consider a standard cosmic scenario, where the inflation era is driven by a scalar field or any other relevant cosmological mechanism. Repulsive PBHs can then be produced from quantum gravity gravitational collapse processes. See here~\cite{1968qtr..conf...87B,Bojowald:2005qw,DeLorenzo:2014pta, Bonanno:2020fgp,Shafiee:2022jfx,Carballo-Rubio:2023mvr,Bambi:2023try,Bonanno:2023rzk,Harada:2025cwd} for relevant works in the literature. We obtain thus in this scenario, a mixture of radiation and PBH matter, with their energy densities scaling respectively as 
$\rho_{\rm rad}=\rho_\mathrm{rad,e}\left(\frac{a_\mathrm{e}}{a}\right)^4$ and
$\rho_{\rm PBH}= \rho_{\mathrm{PBH,e}}\left(\frac{a_\mathrm{e}}{a}\right)^n$.
In the expressions above, $n$ accounts for the non-standard equation-of-state (EoS) of repulsive PBHs and/or for cosmological coupling~\cite{Cadoni:2023lum,Cadoni:2023lqe} whereas $\rho_{\rm PBH}$ denotes a theoretically consistent definition for the  repulsive PBH energy density.

Depending on the parameter $n$ and the the initial PBH abundance, PBHs will ultimately dominate or not the Universe energy. To get thus an insight on this question, one can trivially show that the PBH abundance will scale as
\beq\label{eq:Omega_PBH}
\Omega_\mathrm{PBH} \equiv \frac{\rho_\mathrm{PBH}}{\rho_\mathrm{rad}} = \Omega_\mathrm{PBH,f} \left(\frac{a}{a_\mathrm{f}}\right)^{4-n},
\eeq 
where $\Omega_\mathrm{PBH,f}$ stands for the initial abundance of PBHs at formation.

Interestingly enough, for $n<4$ and for high enough $\Omega_\mathrm{PBH,f}$, PBHs will eventually dominate the energy budget of the Universe as one can see from \Eq{eq:Omega_PBH}. In this case, the cosmic expansion dynamics will be governed by \Eq{eq:H_2_Hayward} with an early era of cosmic acceleration taking place before recombination, often quoted as early dark energy (EDE) era. 
Nevertheless, a PBH-driven EDE epoch before CMB emission with $\Omega_\mathrm{PBH} = \Omega_\mathrm{EDE} > 0.5$ is observationally excluded since it will alter substantially the position of the CMB peaks~\cite{Khlopov:1985fch} suppressing as well the LSS growth rate~\cite{Ferreira:1997au,Ferreira:1997hj,Doran:2001rw}.

However, if $\Omega_\mathrm{PBH,f}$ is small, PBHs will never dominate, with $\Omega_\mathrm{PBH}<0.5$. In this regime, one gets a co-existence of radiation and matter in form of PBHs. Such a mixture of different energy density components requires an extended ``Swiss-Cheese'' inhomogeneous framework~\cite{Carrera:2008pi,Celerier:2024dvs}, which is currently missing to the best of our knowledge. We need to highlight nonetheless that if PBHs with masses around $10^{12}\mathrm{g}$, starting to evaporate slightly before matter-radiation equality, have abundances at that time within the range $0.107<\Omega_\mathrm{PBH}<0.5$, one inevitably produces the correct amount of EDE, compatible with current observational CMB~\cite{Pettorino:2013ia} and LSS~\cite{Smith:2020rxx} EDE constraints being recast as 
\beq
\Omega_\mathrm{EDE}(t_\mathrm{LS})<0.015\quad\mathrm{and}\quad 0.015<\Omega_\mathrm{EDE}(t_\mathrm{eq})<0.107,
\eeq
where $t_\mathrm{LS}$ and $t_\mathrm{eq}$ stand respectively for the times at the last-scattering and matter-radiation equality. Remarkably, we do not have to introduce additional scalar fields playing the role of EDE~\cite{Poulin:2023lkg} since we are met naturally with a PBH-inspired EDE component. An interesting characteristic of this PBH-EDE component is the fact that it decays faster than radiation due to Hawking evaporation, something which is necessary so as to get an increased value of the $H$ parameter at early times. Our PBH-inspired EDE proposed mechanism can then be compatible with the late-time SNIa observations~\cite{Poulin:2018cxd,Kumar:2024soe,Shah:2023sna}, alleviating thus naturally the $H_0$ tension.

\section{Conclusions}

We recap in this work a novel inflationary mechanism driven by repulsive ultra-light PBHs with masses $ m < 5 \times 10^8\,\mathrm{g}$ within the “Swiss-Cheese” cosmological framework. Interestingly, matching such repulsive black hole spacetimes to a homogeneous and isotropic Universe we got an early quasi-de Sitter inflationary era, ending naturally due PBH Hawking evaporation or below a critical energy scale depending on the PBH metric parameters.

Furthermore, this generic PBH-driven cosmic acceleration mechanism can account as well for an EDE component prior to CMB emission. More specifically, PBHs with $ m \sim 10^{12}\,\mathrm{g} $ and abundances $ 0.107 < \Omega^\mathrm{eq}_\mathrm{PBH} < 0.5 $ just before matter-radiation equality can produce the correct amount of EDE necessary to account for the $H_0$ cosmic tension.

\begin{figure}[h!]
\centering
\includegraphics[width=0.8\textwidth]{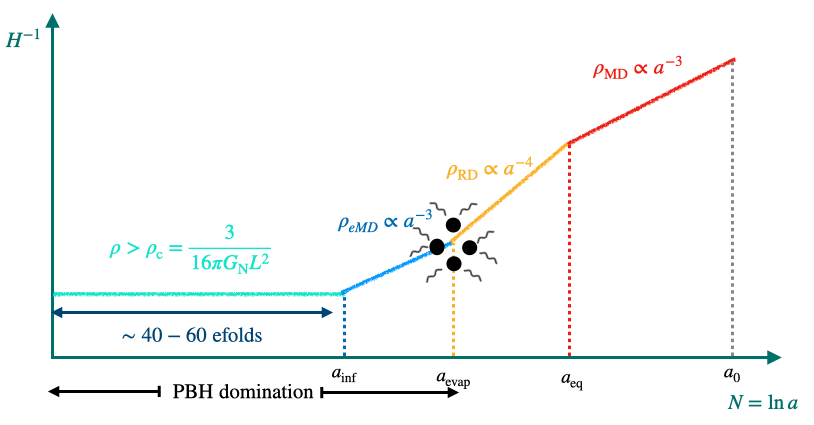}
\caption{{The dynamics of the cosmological horizon $H^{-1}$ for a Universe dominated by repulsive primordial black holes of Hayward type.}}
\label{fig:cosmic_history}
\end{figure}

This alternative cosmic acceleration scenario is explained schematically with \Fig{fig:cosmic_history} where we show the dynamical behaviour of the cosmological horizon $H^{-1}$ as a function of the e-fold number $N$ for a Universe dominated by  regular PBHs of Hayward type with repulsive behaviour. Similar findings are obtained for the Bardeen, Dymnikova and the Schwarzschild-de Sitter spacetimes as well. 

The new inflationary mechanism introduced in \cite{Dialektopoulos:2025mfz} opens new perspectives to study, in particular to find specific quantum gravity frameworks where such PBHs can be produced and study the associated cosmological phenomenology.  Indicatively, we need to mention that such PBH-driven early matter dominated (eMD) eras, like the ones considered here, can give rise to early structure formation. This possibility has been recently explored in~\cite{Jedamzik:2010dq, Barenboim:2013gya,Eggemeier:2020zeg,Hidalgo:2022yed,Domenech:2023afs} and is associated as well with a very rich gravitational wave (GW) phenomenology~\cite{Dalianis:2020gup,Fernandez:2023ddy,Dalianis:2024kjr}. 

Finally, with regard to late-time cosmology, one needs to emphasize that this mechanism may also be relevant for late-time cosmic acceleration, accounting for the so called dark energy problem. This late-time application of the ``Swiss-Cheese" cosmological framework was recently studied in~\cite{Kofinas:2017gfv, Anagnostopoulos:2018jdq} considering Schwarzschild-de Sitter black holes within asymptotic safe gravity.

\begin{acknowledgments}
Theodoros Papanikolaou acknowledges the contribution of the COST Actions CA21136 ``Addressing observational tensions in cosmology with systematics and fundamental physics (CosmoVerse)'' and ``CA23130 - Bridging high and low energies in search of quantum gravity (BridgeQG)'', the LISA Cosmology Working Group as well as the support of the INFN Sezione di Napoli \textit{iniziativa specifica} QGSKY. 
\end{acknowledgments}

\bibliographystyle{JHEP} 
\bibliography{references}

\end{document}

%% file: newcommands.tex
\let\oldsqrt\sqrt
\def\sqrt{\mathpalette\DHLhksqrt}
\def\DHLhksqrt#1#2{%
\setbox0=\hbox{$#1\oldsqrt{#2\,}$}\dimen0=\ht0
\advance\dimen0-0.2\ht0
\setbox2=\hbox{\vrule height\ht0 depth -\dimen0}%
{\box0\lower0.4pt\box2}}

\newcommand{\sss}[1]{{\scriptscriptstyle{#1}}}

\newcommand{\uPl}{\mathrm{Pl}}

\newcommand{\usssPl}{\sss{\uPl}}

\newcommand{\Mp}{M_\usssPl}

\newcommand{\beq}{\begin{equation}}
\newcommand{\eeq}{\end{equation}}
\newcommand{\bea}{\begin{equation}\begin{aligned}}
\newcommand{\eea}{\end{aligned}\end{equation}}

\newlength{\wsingfig}
\newlength{\wdblefig}
\newlength{\wquadfig}
\newlength{\wtriplefig}
\newcommand{\Eq}[1]{Eq.~(\ref{#1})}

\newcommand{\Fig}[1]{Fig.~{\ref{#1}}}